# Detection of Exoplanets Using the Transit Method

*Dennis Afanasev, The George Washington University[\*], Washington, DC 20052*

*dennisafa@gwu.edu*

## Abstract

*I conducted differential photometry on a star GSC 3281-0800, a known host to exoplanet HAT-P-32b, using analysis software AstroImageJ. I plotted the measurements from a series of images taken during the transit, via ADU count given from an earth-based digital CCD camera. I was able to establish a definite light curve and learn more about the properties of this exoplanet.*

---

[\*] *Columbian College of Arts and Sciences, 2020*

## Introduction

Exoplanets are the planets found outside of the solar system. Since the first exoplanet was discovered in 1992, a number of methods for detection have been established. Specific to this paper, I will be evaluating the star known as HAT-P-32 (also catalogued as GSC 3281-0800) and its exoplanet HAT-P-32b using the transit method.

When a planet passes in front of a star, the brightness of that star as observed by us becomes dimmer; which depends on the size of the planet. The data we observe will show a dip in flux if a planet is transiting the star we are observing, as shown in Fig. 1[1].

The figure shows an ideal light curve. In a real world light-curve, the data appears more mangled and with several systematic uncertainties needed to be subtracted. In this research project I will be showing the transit method way of detecting exoplanets. I will be using a free software tool named AstroImageJ [2] in order to conduct my research and obtain the necessary data to demonstrate an observed light curve from raw images.

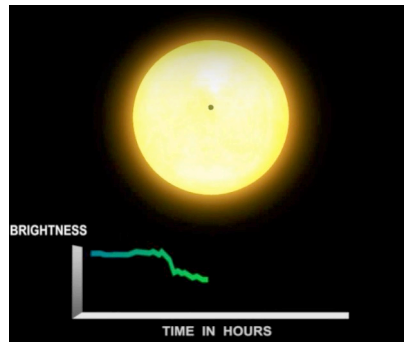

Figure 1. Light curve when planet passes in front of a star.

## Transit Method

I have conducted my research on detecting exoplanets with the transit method. This method relies on taking the light flux of the target star and comparing these values to other stars in the same patch of sky. The transit method is based upon several different processes that will be described throughout this paper. The most useful part of the transit method is that it may be done with a digital CCD (Charged Couple Device) earth-based camera. Based upon NASA's predicted exoplanet transit periods, a camera was set up and aimed at the target stars region of the sky and pictures before, after, and during an exoplanets predicted transit.

These images are then used to generate a dataset of flux values for every frame taken during the session. A downside to the transit method is that it is only effective when observing so called "hot Jupiters," which are planets that are large and close to the star during orbit, and pass directly in front of our field of view from earth.

## Differential photometry

The transit method relies on differential photometry. The idea of differential photometry is to analyse our target star's flux during our imaging session and compare it to other stars in our target star's region of the sky. Specifically, we are looking at ADU counts per pixel. An ADU (Analog to Digital Unit) count is the qualitative value of each CCD's charge output. Each pixel is impacted by a number of photons, which counts the charge and outputs it into a value. A large charge means more photons have hit the CCD pixel. Therefore, we may observe the flux from these images. However, a number of systematic uncertainties must be subtracted from the raw images obtained during the imaging session in order to accurately compare flux changes. Our data lies in the light frames, so we must calibrate the stack of images. Specifically, there are three different types of frames that need to be taken out in order to optimize our raw images. These are the bias, darks, and flat frames. Bias frames are caused by the CCD electronic signal readout during shots, dark frames are simply dark shots which can be obtained by covering the camera lens, and flat frames account for any dust smudges or inconsistencies on the camera lens itself.

## Aperture Photometry

Before obtaining a dataset from the calibrated images, we must calibrate the aperture of the target and comparison stars. The purpose of this is to adjust any uncertainties located in the aperture region, such as background sky noise. The aperture region is the bright outer rings of the actual star itself. It varies in size depending on the focal length of the camera. Typically, the aperture should be set within 20 pixels of the star, and its outer region (called the annulus) must be set to 30-40 pixels. Taking the annulus's values, they are set to the apertures region, thus cancelling out any underlying uncertainties that may affect our data output. Fig 3 shows how this looks in AstroImageJ.

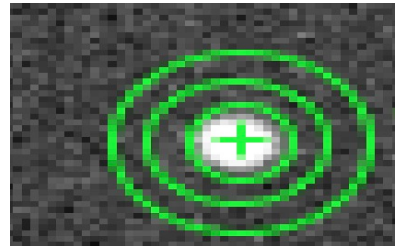

Fig. 3. The second ring shows the aperture region, while the outermost ring is the calibration area.

## Analysis of HAP-P-32

I analysed 758 raw images of the exoplanet HAT-P-32b predicted transit orbit on 2016/01/17 (center of transit as predicted by NASA[3] in Julian Date: 2457404.65827). Using AstroImageJ, I am able to calibrate the images, and analyse the flux of the star HAP-P-32 and fit a model onto the data.

## Calibration

I obtained the necessary flat, bias, and dark frames used to negate the systematic uncertainties from my images. These are frames that must be negated from the images to exclude uncertainties. Using AstroImageJ's image data processor, I input my image directories and calibrated all 758 images. These calibrated images are then saved to a new directory, separate from the original images. Then, they are arranged into a stack (from first to last) and input as a sequence into the program, resulting in Figure 4.

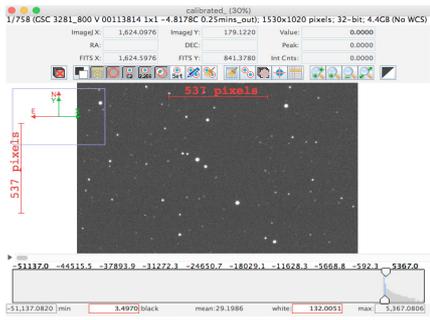

**Figure 4. All calibrated images form a stack**

These images are ready for differential photometry, once the aperture region is defined.

## Defining the Aperture and Differential Photometry

Now that the calibrated images are loaded into the program, the next step is defining the aperture region of our target and comparison stars. I set the apertures pixel range to my best estimate, resulting in 12 pixels for the star, 15 pixels for the aperture region, and about 35 pixels for the annulus. The software will now calibrate the aperture by setting it to the annulus's ADU counts, to eliminate any uncertainties. Note that AstroImageJ automatically eliminates any faint stars in the annulus's region. Once this is done, I set the target stars and comparison stars. I selected several comparison stars and one target star, resulting in Figure 5. Hitting enter evaluates each image in sequence and plots the flux, resulting in Figure 6.

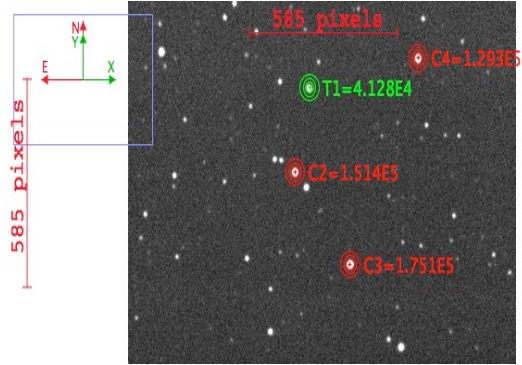

**Figure 5. Selected target and comparison stars**

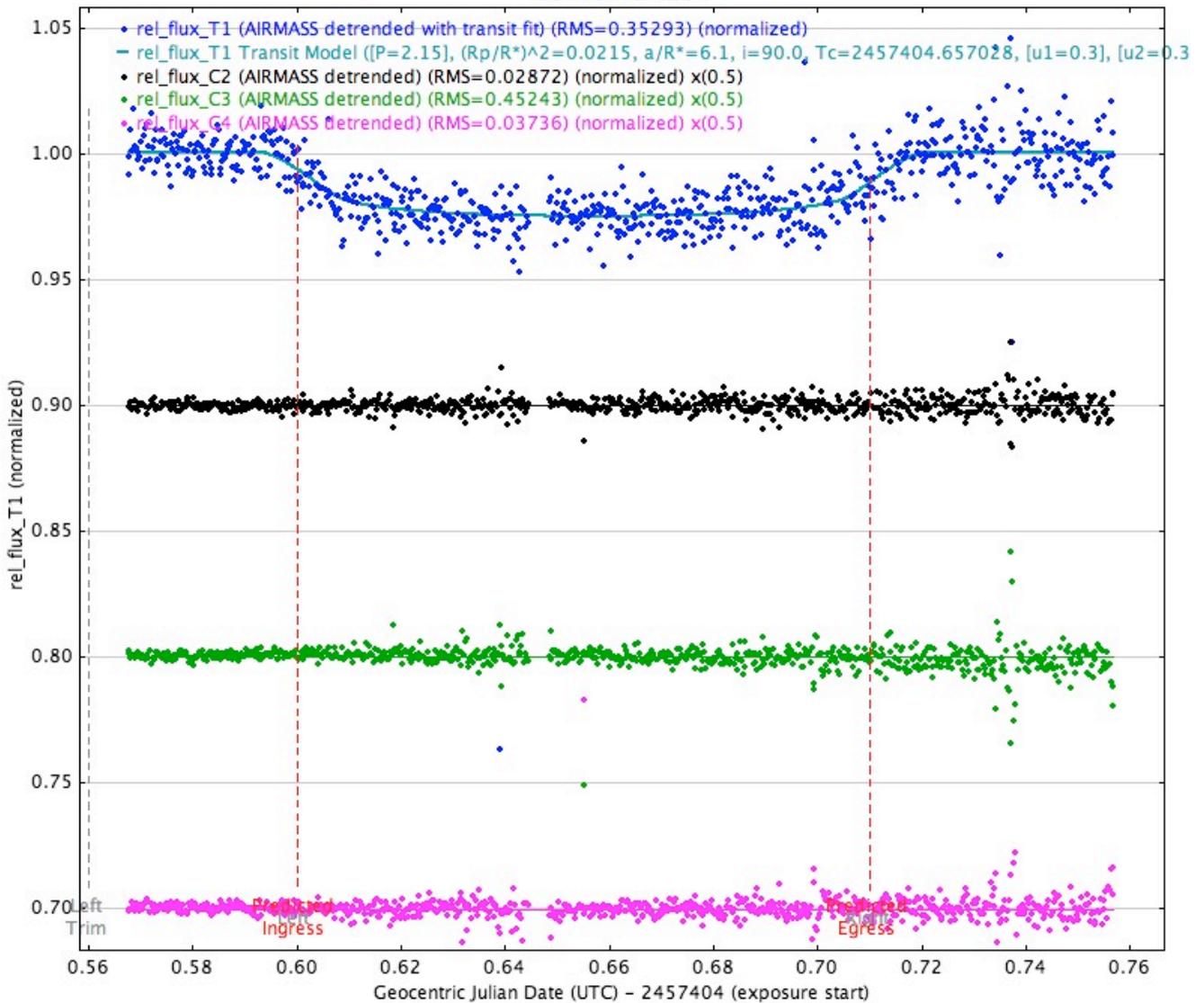

Figure 6. Resulting plots of HAT-P-32b and 3 comparison stars

## Properly plotting the data

It is necessary to specify the proper graph format in which the measurements will be displayed in the AstroImageJ plot GUI (Graphic User Interface.) The beginning date of the observation is recorded in the Julian Date format and is plotted along the X-axis, which is set to its representation of hours and minutes (2 decimal places.) The flux is plotted along the Y-axis. Each star's flux is normalized to points along the Y-axis of my choosing, for visualization purposes. Airmass is a set of data recorded along with the transit photo-shoot. It determines atmospheric conditions and negates them from the data. It is considered an uncertainty.

## Analysis of the plotted data

As seen on Figure 6, our target star (labeled as rel_fluxT1) encounters a 2.1% dip in flux as the exoplanet transits across. Notice the dispersion of the points in the target star in relation to the comparison stars. This is due to the possibility of a binary planetary system or random stellar fluctuations. However, it may not fully be known exactly what the uncertainties are, but we may still draw conclusions from the data regardless. Notice a gap in the data at around 0.65 JD. This is due to some bad images in our calibrated stack, which had to be removed, perhaps because of faulty camera movement or other unknown technical errors. Note that the other stars flux remains relatively linear, essentially telling us that a large body must have passed in front of this star.

## Transit Fit

AstroImageJ is able to automatically fit a model onto the data points, from which we may learn a number of things about the exoplanet. Firstly, the host star parameters must be specified. In particular, the radius (as Rsun), mass (as Msun), and spectral type (F5V for HAP-P-32). From this information, a series of parameters are calculated from the shape of the light curve, such as time in transit and impact angle.  These formulas were obtained from a study conducted by Seager and Mallen-Ornelas [iv] . Some assumptions are made for these formulas; the exoplanets orbit is circular, and the light is from a single star. Periodicity of the planets transit is input manually. For HAT-P-32 it is 2.15 days. The following calculations use variables given from the input below.

$(R_p/R^*)^2$ : The transit depth (which can be expressed as a percentage) derived from:

$$\Delta F \equiv \frac{F_{\text{no transit}} - F_{\text{transit}}}{F_{\text{no transit}}} = \left(\frac{R_p}{R_*}\right)^2$$

Where Rp is the radius of the exoplanet, and R* is the radius of the host star. A result of 0.02145 is obtained from this exoplanet. Multiplied by 100 to get the percentage, it is 2.1% which is consistent with our light curve.

A/R*: The semi-major axis in terms of the host stars stellar radii, given by:

$$\frac{a}{R_*} = \left\{\frac{(1+\sqrt{\Delta F})^2 - b^2[1-\sin^2(t_T\pi/P)]}{\sin^2(t_T\pi/P)}\right\}^{1/2}$$

Where $t_T$ is the total transit duration, P is the period in days, and ΔF is the transit depth, and b is the impact angle of the transit. An amount in value of the host stars radius of 6.0993 is obtained for this respective exoplanet.

Radius of the planet $R_p$ is calculated from:

$$\frac{R_p}{R_\odot} = \frac{R_*}{R_\odot}\sqrt{\Delta F} = \left(k^{1/x}\frac{\rho_*}{\rho_\odot}\right)^{x/(1-3x)}\sqrt{\Delta F}$$

Where k=1 and x=0.8 for main sequence stars.

The final calculated value is expressed in terms of Jupiter radii. A result of 1.85 Rj is obtained for HAT-P-32.

Impact angle (i): Impact angle of the transit with respect to our view from earth, which is derived from:

$$i = \cos^{-1}\left(b\frac{R_*}{a}\right)$$

An impact angle of 89.962 degrees is recorded for HAT-P-32b, which demonstrates a nearly straight transit across the star from our point of view.

## Statistical Aspect

A $\chi^2$ value is automatically calculated by the AstroImageJ software. For our target stars model fit, it is a value of 1630.1119, obtained by:

$$\chi_c^2 = \sum \frac{(O_i - E_i)^2}{E_i}$$

Divided by the number of data points 752 (our d.o.f.) we obtain a value of 2.17. This is not a perfect fit for the data. This is caused by the dispersion problem we saw earlier, which may possibly be due to the host stars random atmospheric activity or an unaccounted binary planet system.

## Summary

HAP-32-B is a close orbiting, "hot Jupiter" planet. Further research would need to be conducted to establish a possible binary system, but for now it is assumed that the data point dispersion is due to unknown stellar fluctuations. The observations I conducted come within close range to a separate study published in the Astrophysical Journal[v]. Although the $\chi^2$ fit is not perfect (with $\chi^2$/d.o.f.=2.17), these results can be justified by the linearity of the comparison stars' flux.

## Acknowledgements

I acknowledge useful advice of Professor Conti of Purdue University on the use of AstroImageJ software and for the access to the image database. This work was a part of an undergraduate research project conducted by Sylvain Guiriec.

# References

[1] Transit Light Curve Tutorial, www.cfa.harvard.edu/~avanderb/tutorial/tutorial.html.

[2] K. Collins, J. Kielkopf, *AstroImageJ: ImageJ for Astronomy*, Astrophysics Source Code Library, ascl:1309.001, http://adsabs.harvard.edu/abs/2013ascl.soft09001C

[3] NASA Exoplanet Archive, exoplanetarchive.ipac.caltech.edu/cgi-bin/TransitView/nph-visibletbls?dataset=transits&sname=HAT-P-32%2Bb&getParams.

[iv] Seager, S., & Mallen-Ornelas, G. (2003). A Unique Solution of Planet and Star Parameters from an Extrasolar Planet Transit Light Curve. The Astrophysical Journal, 585(2), 1038-1055. doi:10.1086/346105

[v] Hartman, J.D. "HAT-P-32b AND HAT-P-33b: TWO HIGHLY INFLATED HOT JUPITERS TRANSITING HIGH-JITTER STARS." Iopscience, 3 Nov. 2011, iopscience.iop.org/article/10.1088/0004-637X/742/1/59/pdf.